\documentclass[mathleft]{an}
\usepackage{graphicx}
\usepackage{txfonts}
\usepackage{times}
\usepackage{natbib}
\bibpunct{(}{)}{;}{a}{}{,}
\begin{document}

\Pagespan{999}{}
\Yearpublication{2012}%
\Yearsubmission{2012}%
\Month{99}%
\Volume{999}%
\Issue{99}%

\title{The puzzle of combination frequencies found in heat-driven pulsators}

\author{P.~I.~P\'{a}pics\inst{1}}
\titlerunning{The puzzle of combination frequencies found in heat-driven pulsators}
\authorrunning{P.~I.~P\'{a}pics}
\institute{Instituut voor Sterrenkunde, KU Leuven, Celestijnenlaan 200D, B-3001 Leuven, Belgium}

\received{15 Aug 2012}
\accepted{?? ??? ????}
\publonline{later}


\abstract{Searching for combinations in the frequency spectra of variable stars is a commonly used method within the asteroseismological community, as harmonics and linear combinations of individual frequencies are expected to appear not only by chance, but also as a characteristic signature linked to different physical phenomena, e.g., nonlinear oscillations, binarity, and rotation. Furthermore it is very important to identify independent frequencies for modelling purposes. We show that using high precision data sets delivered by recent space missions, the distinction between combinations having a real physical meaning and spurious combinations which only appear by chance gets more and more difficult. We demonstrate how careful one should be with the interpretation of such combination frequencies by presenting the statistical distributions derived from artificial data sets. Based on comparisons to observations, we find that, despite the high statistical probability of finding combinations in random data sets (having similar properties to the ones of real stars), there is a significant difference in the number of the lowest order combinations between stars with and starts without real combination frequencies. This way, a search for frequency combinations is very useful when interpreted properly, and when results are compared to simulations.}

\maketitle

\section{Introduction}
\subsection{Frequency analysis of variable stars}
The light (and radial velocity) variations of pulsating stars -- especially of heat-driven oscillators \citep[for a full overview of pulsating stars and asteroseismology, see][]{2010aste.book.....A} which pulsate in a regular, periodic pattern -- can be easily described with a model constructed from a set of sine functions, each of which having well defined parameters. Using an iterative prewhitening procedure to recover the individual modes is a commonly used practice \citep[for a detailed description of the method and discussion on different significance criteria see, e.g.,][]{2009A&A...506..111D, 2012A&A...542A..55P}. This process results in a list of amplitudes ($A_j$), frequencies ($f_j$), and phases ($\theta_j$), by which the light curve can be modelled via $n_f$ frequencies in the well-known form of \[F(t_i)=c+\sum_{j=1}^{n_f}A_j\sin[2\pi(f_j t_i + \theta_j)].\] 

\subsection{Combination frequencies}

If a selected frequency $f_j$ from the complete set of observed frequencies can be written as \[f_j=\sum_{i \neq j = 1}^{n_f} c_{i}f_{i} \pm \Delta f = c_1f_1+c_2f_2+\dots+c_{n_f} f_{n_f} \pm \frac{1}{T}\mathrm{,}\] i.e., as a linear combination of independent $f_i$ frequencies present in the power spectrum (allowing only a small $\Delta f$ deviation from the exact combination value according to the frequency resolution, with $T$ being the timebase of the full dataset, and where the individual coefficients are $c_i\in\mathbb{Z}$), then we term $f_j$ a combination frequency. We define \[\mathcal{O}_{f_j}=\sum_{i \neq j} \left|c_{i}\right|\] to be the combination order, and using this notation, the combination order tells us if we are dealing with independent $(\mathcal{O}_{f_j}=0)$, indistinguishable $(\mathcal{O}_{f_j}=1)$, or combination $(\mathcal{O}_{f_j}\geq 2)$ frequencies. In the latter case, $f_i$ are termed parent frequencies when the corresponding coefficients $c_{i}\neq 0$. Harmonics are a subset of combinations where only $c_{i=k\neq j}\geq 2$ while every other coefficient $c_{i\neq k\neq j} = 0$.

In recent years, space-based instruments like CoRoT \citep{2009A&A...506..411A} and \textit{Kepler} \citep{2010PASP..122..131G} opened up the era of almost uninterrupted data with a precision typically two orders of magnitude better than ground-based photometry, enabling us to study the power spectrum of pulsations without serious aliasing problems combined with a much lower noise level than before. This results in an at least an order of magnitude increase in the number of significant frequencies found in the power spectrum of pulsating stars. This increase in peak density over regions of the periodogram gives rise to ambiguous identifications or misinterpretations of combinations and harmonics.


\section{Statistics of combination frequencies}

\subsection{Creating artificial light curves}\label{artificial}
\begin{figure*}
\resizebox{\hsize}{!}{\includegraphics{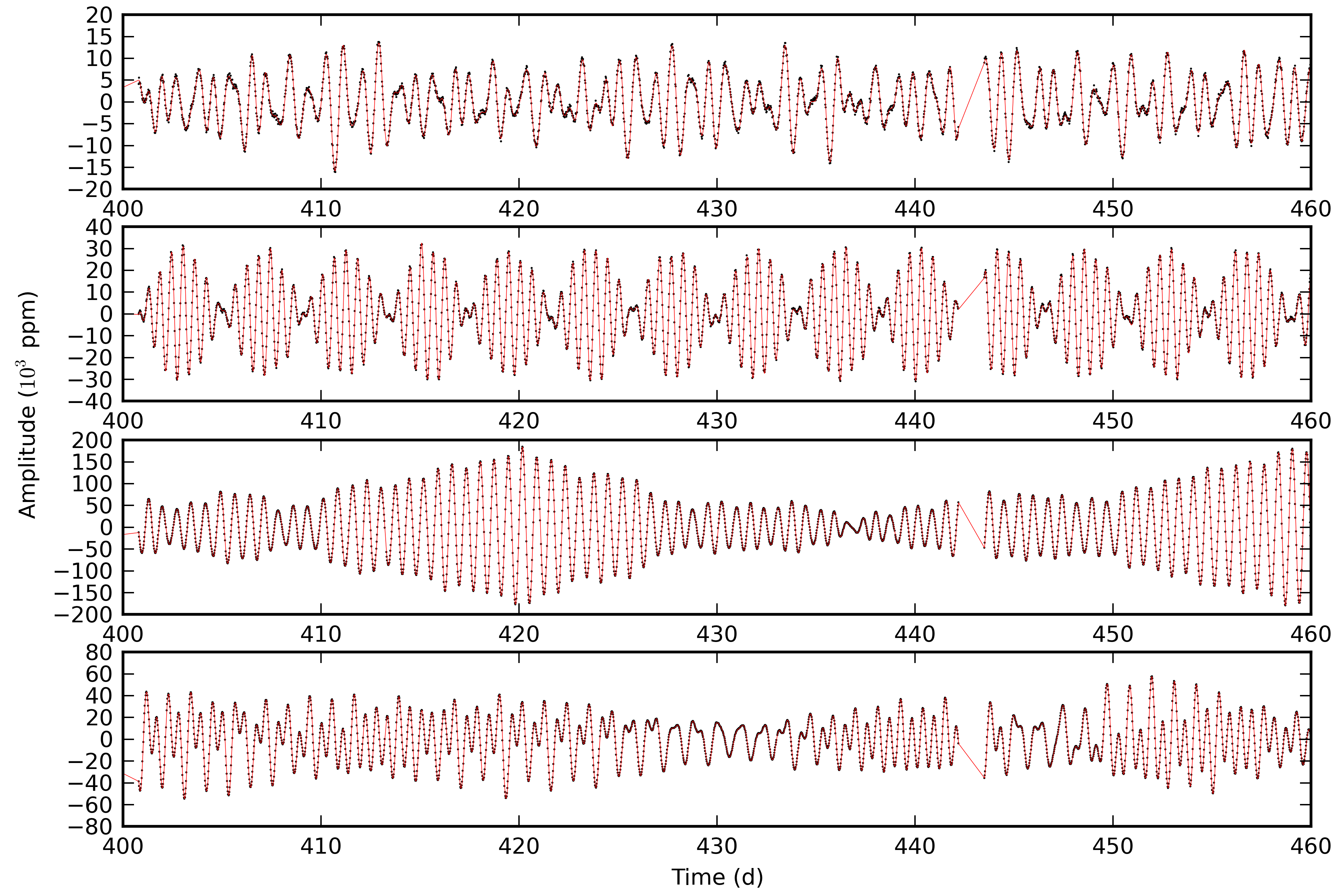}} 
\caption{Example cutout sections of artificial light curves based on the sampling and noise properties of the long cadence data from the \textit{Kepler} satellite and the pulsation properties -- as frequency ranges and amplitude distribution -- of main sequence B stars. The artificial light curves (with added random white noise) are plotted with black dots, while the red solid curves show the corresponding noise-free models constructed from the generated parameter sets.}
\label{artificalexample}
\end{figure*}

It is very important to decide if combination frequencies are real or not to make a proper physical interpretation of the detected signals. Does it make sense to talk about high order combinations in terms of physical phenomena, or, given a massive set of frequencies, will we always find them simply by chance? To test this, we simulated thousands of light curves having a similar but random power spectrum as the main sequence pulsating B stars in our Kepler Guest Observer\footnote{http://keplergo.arc.nasa.gov/} sample (\citeauthor{2012A&A...542A..55P}, in preparation). For each artificial light curve we generated data sets based on a random number of frequencies selected from a typical range of observed real-life frequencies. In addition, we have given these frequencies random amplitudes having a distribution which resembles the observed frequency distribution of our \textit{Kepler} sample. Very realistic light curves emerged (examples can be seen on Fig.\,\ref{artificalexample}), randomly showing beating patterns with both short and long timescales, or features which -- in case of real stars -- we tend to interpret as signs of spotted stars or binarity, and sometimes even quasi-equidistant splittings or frequency series.

\subsection{Searching for combinations}\label{searchinginrandom}
We used a conservative method to look for combination frequencies using the subset of prewhitened frequencies with the fifty highest amplitudes. We set the maximum number of combination terms $c_if_i$ to 3, while allowing $\left|c_i\right|\in\left[0,3\right]$, to avoid extremely high -- and physically unlikely -- combination orders. We assume that combination peaks will have lower amplitudes than their parents, so we only try linear combinations in the subset where $i<j$, and where previously frequency $f_i$ was found to be independent (so we do not use combination peaks as parents). To be as conservative as possible, we use an amplitude criterion requiring the relation \[\frac{\sum A_i}{\mathcal{O}_{f_j}} > A_j\] to be true for peaks accepted as combination frequencies, which -- in an empirical way -- ensures that the $A_i$ amplitudes of the parent peaks are strong enough to create a combination with the given $A_j$ amplitude. 

These tests enabled us to check what is the probability of finding harmonics with a given order $\mathcal{O}$ in a random dataset, just by chance, without forcing the appearance of harmonics during the generation of these simulated light curves. Looking at the distributions from a run of simulations containing one thousand artificial data sets in Fig.\,\ref{comparison_all} we can see that, on average, only $17.3\%$ of the 50 strongest peaks are identified as independent, $4.5\%$ as $\mathcal{O}=2$ combinations, while $78.2\%$ as higher order combinations. The distribution of independent frequencies and $\mathcal{O}=2$ combinations are relatively narrow, so significant deviations will stand out very clearly.

\subsection{Comparison to observations}

\begin{figure*}
\resizebox{\hsize}{!}{\includegraphics{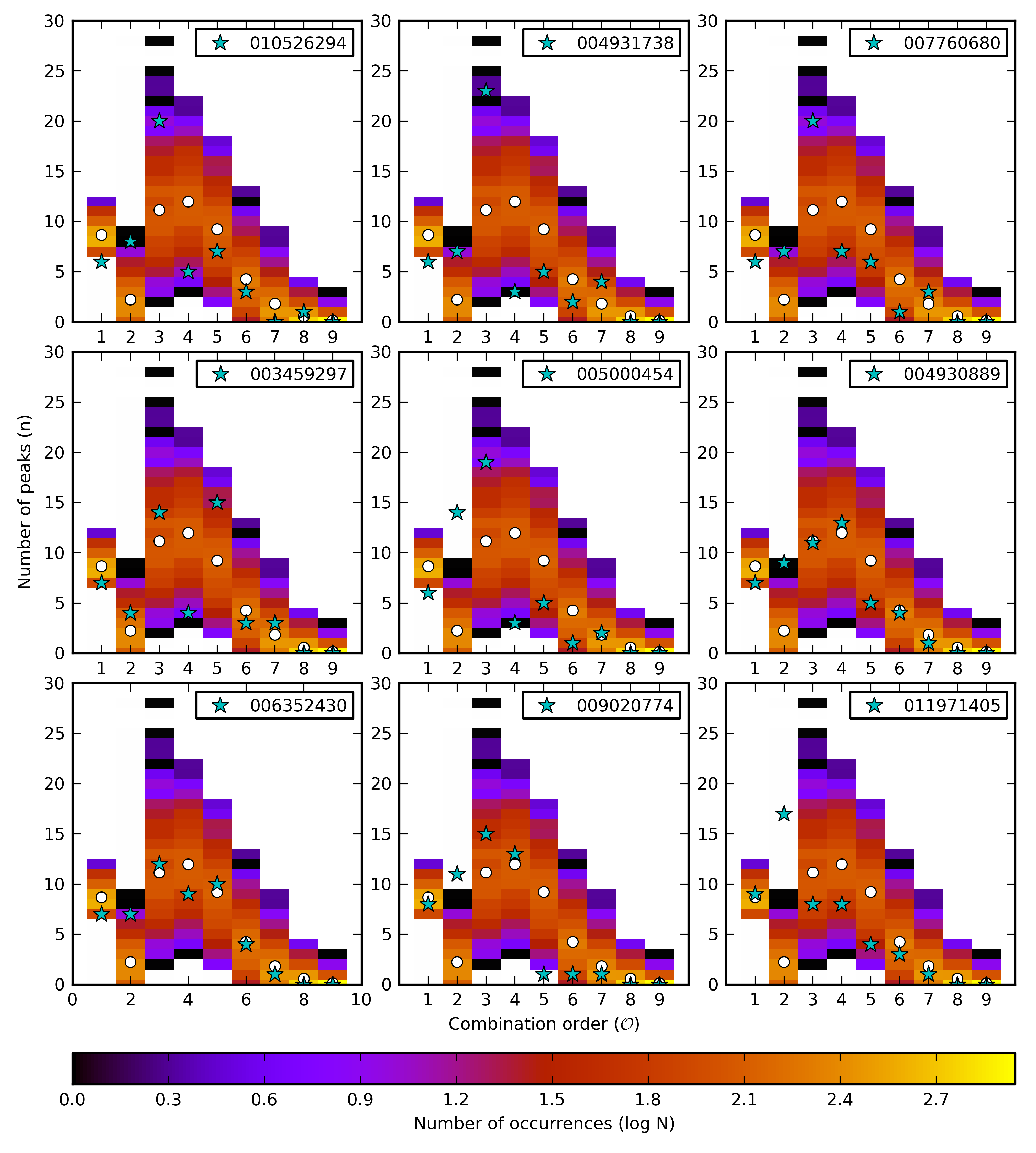}} 
\caption{The observed number of combination frequencies per order for our nine \textit{Kepler} stars (shown in individual subplots with the corresponding KIC numbers, plotted with cyan coloured asterisk symbols) compared to the statistical distributions (plotted as log scale coloured 2D histograms in the background where colours denote the number ($\log N$) of occurrences of having $n$ combination frequencies with a given order ($\mathcal{O}$) in the full set of artificial data sets, and where the average values per order are also  plotted with white circles) from simulations described in Sect.\,\ref{artificial} and Sect.\,\ref{searchinginrandom}.}
\label{comparison_all}
\end{figure*}

\begin{figure}
\resizebox{\hsize}{!}{\includegraphics{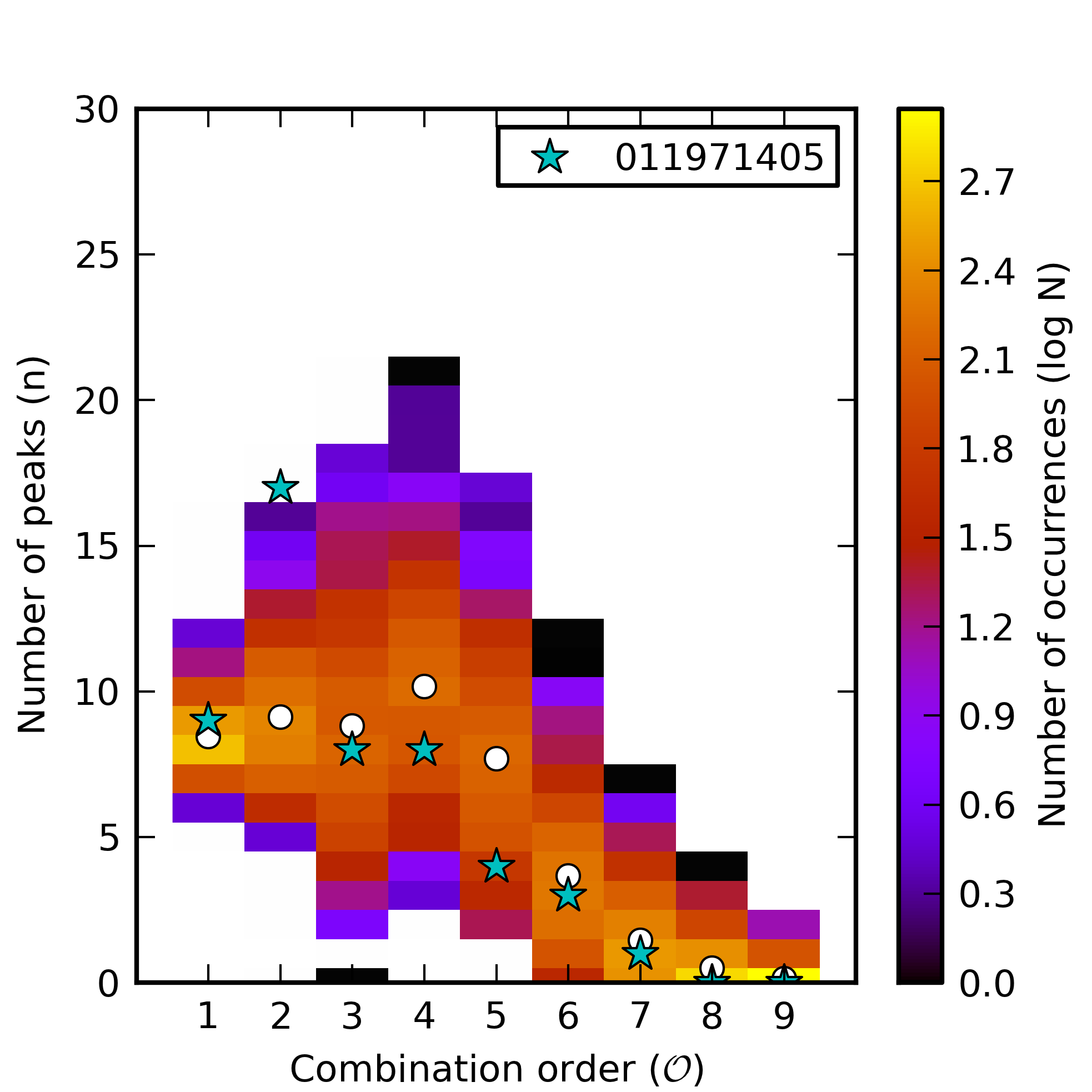}} 
\caption{The same as the \textit{bottom right} subplot in Fig.\,\ref{comparison_all} but now showing the distribution of combination frequencies with a given order when the appearance of $\mathcal{O}=2$ combinations is enforced during the simulations.}
\label{comparison_forced}
\end{figure}

We used the same method outlined in Sect.\,\ref{searchinginrandom} for the frequency spectra of the nine stars of our Guest Observer sample (we analysed Q1-Q6 data for these tests) to see if the observed number of given order combinations fall in the expected distributions (see Fig.\,\ref{comparison_all}). For most of the stars, and most of the combination orders this was the case. Significant deviations can only be seen in three cases and only at combinations having $\mathcal{O}=2$. For these stars the observed number of $\mathcal{O}=2$ combinations is much higher than predicted by the simulations, which means that there has to be a real, physical effect behind the occurrence of combination frequencies. Forcing the occurrence of $\mathcal{O}=2$ combinations in the artificial data sets, we create much better matching distributions for these stars (see Fig.\,\ref{comparison_forced}).

\section{Conclusions}

Seismic modelling of heat-driven oscillators can not be successful using hundreds of frequencies without mode identification constraints, as raising the number of modes which are used in the fitting process quickly raises the complexity beyond the reach of our computing power, or simply beyond the precision of our present-day models.

Filtering out combination frequencies when looking for a set of independent frequencies can be difficult \citep{2011A&A...534A.125U,2011MNRAS.415.1691B}, e.g., the presence of binary components or rotation can give rise to very long series of harmonics \citep{2011MNRAS.413.2651B,2003A&A...407.1029C}, and one can also treat rotationally split multiplets as a kind of combinations if the rotation period is present in the power spectrum \citep{2011MNRAS.414.1721B}. Most of these effects can be correctly interpreted from spectroscopy, or by looking for clear rotationally split multiplets in the power spectrum, so in ideal cases these should be already known when searching for combinations. In case of rotational or binary signature, higher order harmonics are expected to appear, so these should be identified making sure that results from photometry and spectroscopy agree with each other.

After this step it does not make sense to look for high order combinations, as our tests showed that those appear very easily from random sets even when a conservative search method is used. We strongly suggest to look at the observed number of $\mathcal{O}=2$ combinations and compare it to statistics of artificial sets created using frequency ranges and amplitude distributions which represent the related observed quantities well. Only if there is a significant deviation between the observed and predicted number of combinations, a clear signature of an underlying physical effect should be concluded, otherwise any combinations found in the data are most probably mere coincidence. Only when the presence of $\mathcal{O}=2$ combinations is proven one should proceed and try to identify higher order combinations. Additional spectroscopy might be very useful to help deciphering the physical origin of combination frequencies. Ignoring such data can easily lead to meaningless identification and over-interpretation of high order multiplets in variable star research.


\acknowledgements
The research leading to these results has received funding from the European Research Council under the European Community's Seventh Framework Programme (FP7/2007--2013)/ERC grant agreement n$^\circ$227224 (PROSPERITY), as well as from the Belgian Science Policy Office (Belspo, C90309: CoRoT Data Exploitation).

\bibliographystyle{aa}
\bibliography{combinations}

\end{document}